\documentclass{IEEEtran4PSCC}

\ifCLASSINFOpdf
   \usepackage[pdftex]{graphicx}
\else
   \usepackage[dvips]{graphicx}
\fi
\usepackage[cmex10]{amsmath}

\renewcommand{\baselinestretch}{0.944}
\usepackage{multirow}

\makeatletter
\let\old@ps@headings\ps@headings
\let\old@ps@IEEEtitlepagestyle\ps@IEEEtitlepagestyle
\def\psccfooter#1{%
    \def\ps@headings{%
        \old@ps@headings%
        \def\@oddfoot{\strut\hfill#1\hfill\strut}%
        \def\@evenfoot{\strut\hfill#1\hfill\strut}%
    }%
    \def\ps@IEEEtitlepagestyle{%
        \old@ps@IEEEtitlepagestyle%
        \def\@oddfoot{\strut\hfill#1\hfill\strut}%
        \def\@evenfoot{\strut\hfill#1\hfill\strut}%
    }%
    \ps@headings%
}
\makeatother

\psccfooter{%
        \parbox{\textwidth}{\hrulefill \\ \small{22nd Power Systems Computation Conference} \hfill \begin{minipage}{0.2\textwidth}\centering \vspace*{4pt} \includegraphics[scale=0.06]{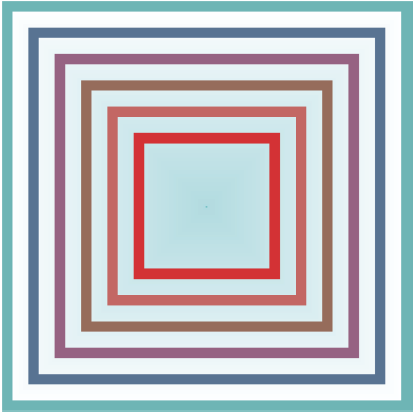}\\\small{PSCC 2022} \end{minipage} \hfill \small{Porto, Portugal --- June 27 -- July 1, 2022}}%
}

\begin{document}

%
\title{Physics-Informed Neural Networks for AC Optimal Power Flow}

\author{
\IEEEauthorblockN{Rahul~Nellikkath, Spyros~Chatzivasileiadis}
\IEEEauthorblockA{Center for Electric Power and Energy \\ Technical University of Denmark (DTU) \\
Kgs. Lyngby, Denmark \\
\{rnelli, spchatz\}@elektro.dtu.dk}
}

\maketitle

\begin{abstract}
This paper introduces, for the first time to our knowledge, physics-informed neural networks to accurately estimate the AC-OPF result and delivers rigorous guarantees about their performance. Power system operators, along with several other actors, are increasingly using Optimal Power Flow (OPF) algorithms for a wide number of applications, including planning and real-time operations. However, in its original form, the AC Optimal Power Flow problem is often challenging to solve as it is non-linear and non-convex. Besides the large number of approximations and relaxations, recent efforts have also been focusing on Machine Learning approaches, especially neural networks. So far, however, these approaches have only partially considered the wide number of physical models available during training. And, more importantly, they have offered no guarantees about potential constraint violations of their output. Our approach (i) introduces the AC power flow equations inside neural network training and (ii) integrates methods that rigorously determine and reduce the worst-case constraint violations across the entire input domain, while maintaining the optimality of the prediction. We demonstrate how physics-informed neural networks achieve higher accuracy and lower constraint violations than standard neural networks, and show how we can further reduce the worst-case violations for all neural networks.
\end{abstract}

\begin{IEEEkeywords}
AC-OPF, Physics-Informed Neural Network, Worst-Case Guarantees.
\end{IEEEkeywords}

\thanksto{\noindent This work is supported by the FLEXGRID project, funded by the European Commission Horizon 2020 program, Grant Agreement No. 863876, and by the ERC Starting Grant VeriPhIED, Grant Agreement No. 949899}

\section{Introduction}
Power system operators, electricity market operators, and several other actors increasingly use Optimal Power Flow (OPF) for the planning and real-time operation of power systems. Countless instances of OPF need to be solved when it comes to evaluating uncertain scenarios, finding optimal control setpoints, bidding strategies, or determining the electricity market clearing. However, the exact formulation of the AC-Power Flow equations in the OPF problem is non-linear and non-convex \cite{Convex}, which could result in significant difficulties for convergence and long computational times. A large body of literature exists on deriving approximations of the AC-OPF problem, with the most popular ones being the linearized DC-OPF, other linear or convex approximmations, and convex relaxations, that will ensure fast computation speed and convergence guarantees \cite{DCOPF}.

Most recently, there is a revived interest in machine learning methods to accurately estimate the AC-OPF solution \cite{MLDCOPF} \cite{recent}, which have demonstrated a computation speedup of 100-1'000 times compared with the conventional methods. This means that in the time it would take to assess one scenario by solving one AC-OPF instance, we can now assess up to 1'000 scenarios simultaneously. However, these machine learning algorithms experience two significant challenges. First, the availability and quality of training datasets: to train a neural network with considerable accuracy, we need OPF results for a huge set of operating points that will cover both normal and abnormal situations. Such datasets often do not exist, or it is often challenging to generate. Convex relaxation techniques were proposed in \cite{Database, venzke2019efficient} to efficiently generating such large datasets, concentrating closer to the security boundary. Along the lines of improving the performance of such Machine Learning algorithms, a method was proposed in \cite{Andreas_verify} to identify and include adversarial examples in the training data set during training.  

The second major issue limiting the Neural Network (NN) widespread adaptation is that, so far, none of the proposed machine learning algorithms have supplied any worst-case performance guarantee. With OPF often used for safety-critical applications, the neural network estimates must not violate any OPF constraints, e.g., line, voltage, or generator limits. To mitigate these limitations, the NN predictions can be post-processed to satisfy generation constraints as proposed in \cite{Pred_AC_OPF} \cite{dc3} for AC-OPF. However, this could negatively impact the optimality of the solution. A few methods have also offered to include the constraint violations in the NN loss function \cite{deepopf}, \cite{Pred_AC_OPF}. However, none of these proposed algorithms provide any worst-case performance guarantees for the AC-OPF problem. The only works that have derived worst-case performance guarantees have so far focused only on the DC-OPF problem \cite{Andreas, Rahul}. 

This paper attempts to address both of these challenges by using a physics-informed neural network for AC-OPF applications. First, we introduce the physical equations in the form of the AC-OPF Karush-Kuhn-Tacker (KKT) conditions inside the neural network training. By doing that, the neural network can reduce its dependency on the size and quality of the training dataset, and instead, it can determine its optimal parameters based on the actual equations that it aims to emulate \cite{misyris}. 

Second, we introduce methods that extract worst-case guarantees for generation and line flow constraint violations of the neural network AC-OPF predictions, extending our previous work presented in \cite{Andreas} and \cite{Rahul}. Through that, we (i) determine the worst violations that can result from any neural network output across the whole input domain, and (ii) propose methods to reduce them.

This paper is structured as follows: Section II describes the AC-OPF problem and its KKT formulation, introduces the proposed physics-informed neural network training architecture, and the optimization algorithm used to determine the worst-case guarantees. Section III provides the simulation setup used and delivers the results demonstrating the performance of physics-informed neural networks. Section IV discusses the possible opportunities to improve the system performance and concludes. 

\section{Methodology}
\subsection{AC - Optimal Power Flow}
The AC Optimal Power Flow (AC-OPF) problem for generation cost minimization in a system with $N_b$ number of buses, $N_g$ number of generators and $N_d$ number of loads is a Quadratically Constrained Quadratic Programing (QCQP) problem. The objective function for the cost minimization can be written as follows: 
\begin{equation}
    \underset{\mathbf{P_d,Q_d}}{\mathrm{min}}\text{ }  \mathbf{c}^T_p\mathbf{P}_g + \mathbf{c}^T_q \mathbf{Q}_g \label{obj}
\end{equation}
where vector $\mathbf{c}^T_p$ and $\mathbf{c}^T_q$ refers to the linear cost terms for the active and reactive power generations $\mathbf{P}_g$ and $\mathbf{Q}_g$, respectively. The optimal generation values depend on the active and reactive power demand, denoted by $\mathbf{P}_d$ and $\mathbf{Q}_d$, and the network constraints. For a given demand, the active and reactive power injection at each node $n \in N_b$ can be represented as follows:
\begin{align}
    p_n = p_n^g - &p_n^d  &\forall n\in N_b \label{P_n} \\
    q_n = q_n^g - &q_n^d  &\forall n\in N_b \label{Q_n}
\end{align}
where $p_n$ and $q_n$ denote the active and reactive power injection at bus $n$ and ${p}_n^g$, ${q}_n^g$, ${p}_n^d$ and ${q}_n^d$ specifies the active and reactive power generation and demand, respectively at bus $n$. The power flow equations in the network can be expressed as follows:
\begin{align}    
    p_n = \sum_{k=1}^{N_b} v_n^r(v_k^r G_{nk} - v_k^i B_{nk}) + v_n^i(v_k^i G_{nk} + v_k^r B_{nk}) \\
    q_n = \sum_{k=1}^{N_b} v_n^i(v_k^r G_{nk} - v_k^i B_{nk}) - v_n^r(v_k^i G_{nk} + v_k^r B_{nk})
\end{align}
where real and imaginary part of the voltage at bus $n$ is given by $v^r_n$ and $v^i_n$ and, the conductance and susceptance of the line $nk$ is denoted by $G_{nk}$ and $B_{nk}$.  If the real and reactive parts of voltage are combined into a vector of size $2N_b \text{x} 1$ as $\mathbf v = [(\mathbf v^r)^T , (\mathbf v^i)^T]^T$. Then, the power flow equation can be simplified as follows \cite{psse}:
\begin{align}
    \mathbf{v}^T \mathbf{M}_{p}^n \mathbf{v} = &p_n        & \forall n\in N_b \label{PF_p}\\
    \mathbf{v}^T \mathbf{M}_{q}^n \mathbf{v} = &q_n            & \forall n\in N_b \label{PF_q}
\end{align}
Other than the power flow equations the optimal generation values should also satisfy the active and reactive power generation limits.
\begin{align}
    \underline{p}_n^g \leq &\mathbf{v}^T \mathbf{M}_{p}^n \mathbf{v}  \leq \overline{p}_n^g & \forall n\in N_g  \label{Pg_lim} \\
    \underline{q}_n^g \leq &\mathbf{v}^T \mathbf{M}_{q}^n \mathbf{v}\leq \overline{q}_n^g & \forall n\in N_g \label{Qg_lim}
\end{align}

Similarly, the voltage and line current flow constraints for the power system can be represented in a matrix form as follows: 

\begin{align}
    \underline{\mathbf v}^n \leq \mathbf{v}^T \mathbf{M}_{v}^n \mathbf{v} &\leq \overline{\mathbf v}^n & \forall n\in N_b \label{V_lim} \\
    {\ell}_{mn} = \mathbf{v}^T \mathbf{M}_{i_{mn}} & \mathbf{v} \leq \overline{\mathbf \ell}_{mn} & \forall mn\in N_l \label{I_lim}
\end{align}
where $\mathbf{M}_{v}^n := e_ne^T_n + e_{N_b+n}e_{N_b+n}^T$ and $e_n$ is a $2N_b \text{x} 1$ unit vector with zeros at all the locations except $n$. The square of magnitude of upper and lower voltage limit is denoted by $\overline{\mathbf v}^n$ and $\underline{\mathbf v}^n$ respectively, and the square of magnitude of line current flow in line $mn$ is represented by ${\ell}_{mn}$ and matrix  $\mathbf{M}_{i}^{mn} = |y_{mn}|^2(e_m - e_n)(e_m -e_n)^T + |y_{mn}|^2(e_{N_b+m} - e_{N_b+n})(e_{N_b+m} - e_{N_b+n})^T$ where $y_{mn}$ is the admittance of branch $nm$. Assuming the slack bus $N_{sb}$ acts as an angle reference for the voltage, we will have:
\begin{equation}
    v^i_{N_{sb}} = \mathbf{v}^T \mathbf{e}_{N_b + N_{sb}} \mathbf{e}^T_{N_b +N_{sb}} \mathbf{v} = 0 
    \label{V_sb}
\end{equation}
The constraints \eqref{P_n}-\eqref{Q_n},\eqref{PF_p}-\eqref{V_sb} and the objective function for the AC-OPF problem \eqref{obj} can be simplified as follows: 
\begin{subequations}
\begin{align}
    \underset{\mathbf{v},\mathbf G}{\mathrm{min}} \text{ } \mathbf{c}^T \mathbf G \label{pri_1}\\
    s.t \text{ } \mathbf{v}^T \mathbf{L}_l \mathbf{v} &= a_l^T \mathbf G +b_l^T \mathbf{D}, & l=1:L &:\lambda_l \label{pri_2}\\
    \mathbf{v}^T \mathbf{M}_m \mathbf{v} &\leq d_m^T\mathbf{D} + f_m, & m = 1 :M &:\mu_m \label{pri_3}
\end{align}
\end{subequations}
where $\mathbf{G} = [\mathbf{P}_g^T , \mathbf{Q}_g^T]^T$, $\mathbf{D} = [\mathbf{P}_d^T , \mathbf{Q}_d^T]^T$, and $\mathbf{c}^T$ is the combined linear cost terms for the active and reactive power generations. Then the equality constraints \eqref{PF_p}-\eqref{PF_q} and \eqref{V_sb} can be represented by the $\mathbf L=2N_b+1$ constraints in \eqref{pri_2}. Similarly, the inequality constraints \eqref{Pg_lim}-\eqref{I_lim} can be represented by the $\mathbf M=4N_g+2N_b+N_l$ constraints in \eqref{pri_3}. The corresponding Lagrange multipliers are denoted by $\lambda_l$ and $\mu_m$. 

The Lagrangian function $\mathcal{L}$ for the AC-OPF can be formulated as follows:
\begin{equation} \label{eq1}
\begin{split}
\mathcal{L}(\mathbf{x},\lambda,\mu, D)  = {c}^T \mathbf G & + \sum_{l=1}^L \lambda_l( \mathbf{v}^T \mathbf{L}_l \mathbf{v} - a_l^T \mathbf G - b_l^T \mathbf{D})\\
 & + \sum_{m=1}^M \mu_m(\mathbf{v}^T \mathbf{M}_m \mathbf{v} - d_m^T\mathbf{D} - f_m)
\end{split}
\end{equation}
where $\mathbf{x}=\{\mathbf G,\mathbf{v}\}$. So, the KKT conditions can be formulated as follows:
\begin{align}
c = \sum_{l=1}^L \lambda_l a_l \label{Stat1} \\
\bigg( \sum_{l=1}^L \lambda_l \mathbf{L}_l + \sum_{m=1}^M \mu_m \mathbf{M}_m  \bigg) &= 0 \label{Stat2}\\
\mu_m(\mathbf{v}^T \mathbf{M}_m \mathbf{v} - d_m^T\mathbf{D} - f_m) &= 0  &m = 1 :M\label{comp} \\
\mu_m &\geq 0 & m = 1 :M \label{dual}\\
\eqref{pri_2}-\eqref{pri_3} \label{Prim}
\end{align}
where the stationarity condition is given in \eqref{Stat1} and \eqref{Stat2}, the complementary slackness condition is given by \eqref{comp} and dual feasibility is given by \eqref{dual}. For an AC-OPF problem these KKT conditions act as a necessary condition for optimality. 
\subsection{Physics Informed Neural Network}\label{SecPINN}
\begin{figure}[htbp]
\centerline{\includegraphics[scale=.4]{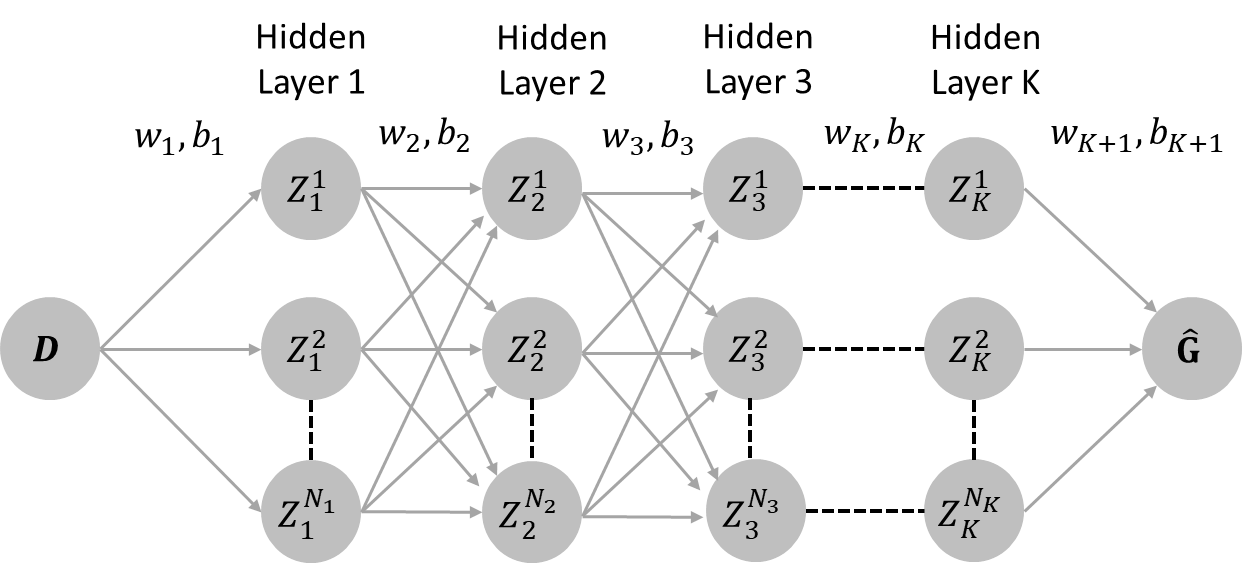}}
\caption{Illustration of the neural network architecture to predict the optimal generation active and reactive power outputs $\mathbf{\hat G}$ using the active and reactive power demand $\mathbf{D}$ as input: There are K hidden layers in the neural network with $N_k$ neurons each. Where k = 1, ...,K.}
\label{NN_basic}
\end{figure}

This section introduces the physics-informed neural network architecture used for predicting the AC-OPF active and reactive power generation setpoints $\mathbf{ G}$, given active and reactive power demand $\mathbf{D}$ as the input. A conventional neural network (NN) is a group of interconnected nodes correlating the input and the output layers, as shown in Fig. \ref{NN_basic}. Nodes connecting the input and output layers will be divided into $K$ number of hidden layers with $N_k$ number of neurons in hidden layer $k$ and the edges connecting the neurons will have a weight $\mathbf w$ and a bias $\mathbf b$ associated with them. Also, each neuron in the neural network will have a nonlinear activation function linked with them.

The output of each layer in the NN can be denoted as follows:
\begin{equation}
    Z_{k+1} = \pi(w_{k+1}Z_k+b_{k+1})
\end{equation}
where $Z_{k+1}$ is the output of layer $k+1$, $w_{k+1}$ and $b_{k+1}$ are the weights and biases connecting layer $k$ and $k+1$. $\pi$ is the nonlinear activation function. We chose the ReLU as the nonlinear activation function in this work, as it is observed to accelerate the neural network training [14]. The ReLU activation function will return the input if the input is positive and return zero if the input is negative or zero. The ReLU activation function can be formulated as follows:
\begin{align}
    \hat Z_{k+1} &= w_{k+1}Z_{k}+b_{k+1}\label{NN1}\\
    Z_{k+1} &= \max( \hat Z_{k+1},0)\label{Relu}
\end{align}

During the NN training, the backpropagation algorithm will modify the weights and biases to predict the optimal generation setpoint for the AC-OPF problem accurately. 

\begin{figure}[htbp]
\centerline{\includegraphics[scale=.38]{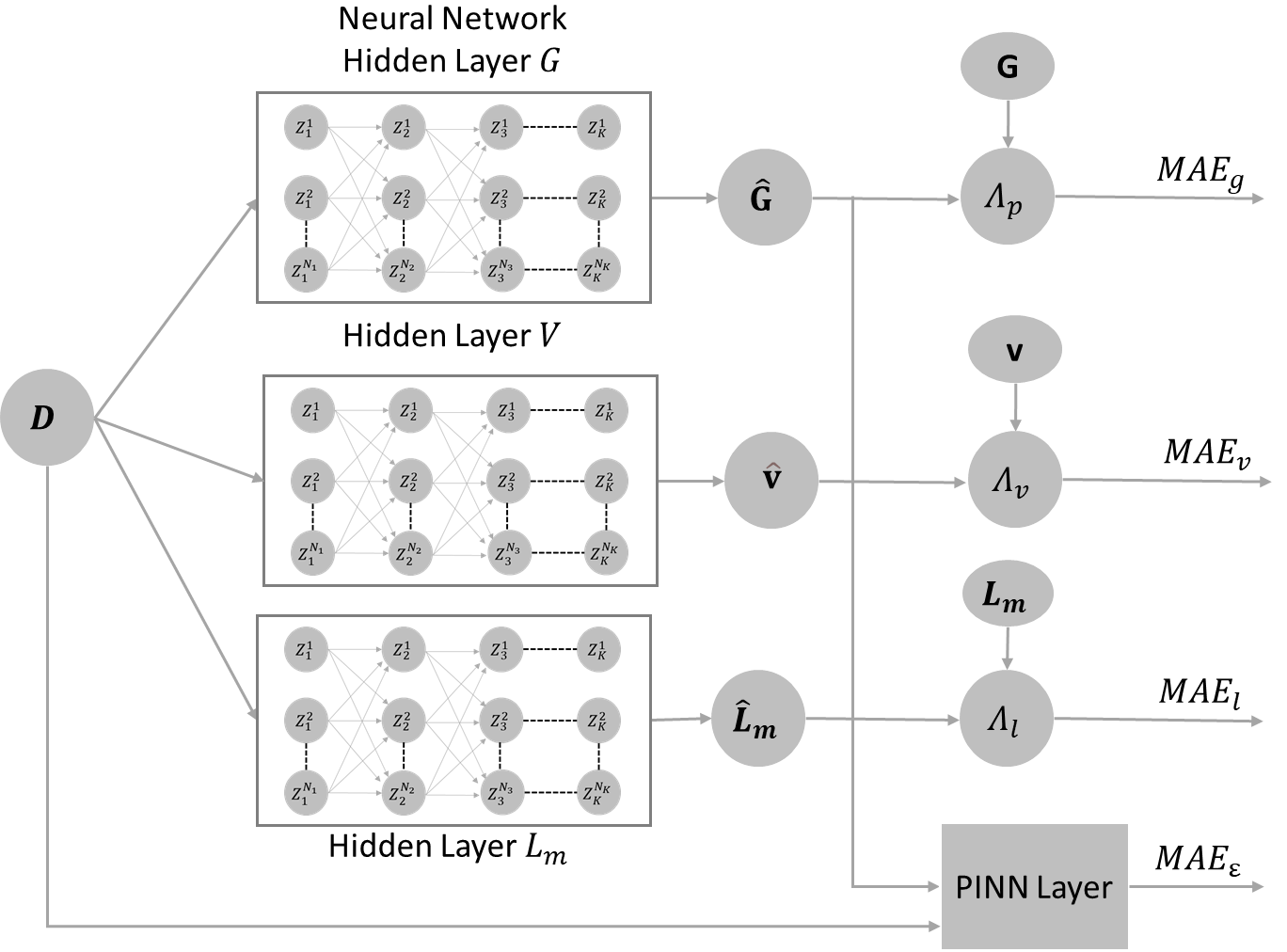}}
\caption{Illustration of the physics-informed neural network architecture to predict the optimal active and reactive power generation outputs $\mathbf{\hat G}$, voltage setpoints $\hat{\mathbf v}$ and dual variables $\mathbf{\hat{L}_m}$ utilizing the active and reactive power demand $\mathbf{D}$ as input. Hidden layers that predict $\mathbf{\hat G}$,  $\hat{\mathbf v}$ and $\mathbf{\hat{L}_m}$ are independent of each other. During training, the weights $\mathbf w$ and biases $\mathbf b$ are modified to minimizes the mean absolute errors $MAE_g$,  $MAE_v$, $MAE_l$ and $MAE_{\epsilon}$}
\label{PINN_Fig}
\end{figure}

In case of a physics-informed neural network, on top of comparing the NN predictions to the AC-OPF setpoints of the training database, the validity of the physical equations governing the problem will also be accessed during NN training (see \cite{raissi_PINN} \cite{misyris}, and our previous work \cite{nellikkath} for DC-OPF applications). Since the optimal value should satisfy the KKT conditions given in \eqref{Stat1} - \eqref{Prim}, the disparities in the KKT condition, denoted by $\epsilon$, as shown in \eqref{EStat}-\eqref{EPrim} are added to the NN training loss function \eqref{MAE} and minimized during training. The proposed physics-informed neural network structure is given in Fig. \ref{PINN_Fig}. 

The voltage values and the dual variables required for calculating the discrepancy in the KKT conditions are predicted using a separate set of hidden layers, as we have observed that this increases the accuracy while maintaining a smaller neural network size. 

The discrepancy in KKT conditions are calculated as follows:
\begin{subequations}
\begin{align}
    \epsilon_{stat} &= \lvert c -\sum \hat{\lambda}_l a_l \lvert + \lvert \bigg( \sum \hat{\lambda}_l\mathbf{L}_l + \sum \hat{\mu}_m \mathbf{M}_m  \bigg)\lvert \label{EStat}\\
    \epsilon_{comp} &=  \sum \lvert \hat{\mu}_m(\mathbf{\hat{v}}^T \mathbf{M}_m \mathbf{\hat{v}} - d_m^T\mathbf{D} - f_m) \lvert \\
    \epsilon_{dual} &= \pi(\hat{\mu}_m)\label{Edual} \\
    \epsilon_{prim} &= \sum \lvert \mathbf{\hat{v}}^T \mathbf{L}_l \mathbf{\hat{v}} - a_l^T \mathbf G - b_l^T \mathbf{D}\lvert \label{EPrim}
\end{align}
\end{subequations}
where $\hat{\mathbf v}$ is the voltage values predicted by the hidden layers $V$ and $\hat{\lambda}_l$ and $\hat{\mu}_m$ are the dual variables predicted by hidden layers $L_m$ in Fig.~\ref{PINN_Fig}. The absolute value of the stationarity condition is measured in $\epsilon_{stat}$, and the error in complementary slackness conditions \eqref{comp} is given by $\epsilon_{com}$. The ReLU activation function, represented by $\pi$ in \eqref{Edual}, is used to measure the dual feasibility violation. If the neural network prediction is the optimal value, then these errors will be zero.

Since we also have the KKT conditions to measure the accuracy of the NN prediction, we can have collocation points in the training set. Like the NN training data points, the collocation points are a set of random input values from the input domain. However, unlike the training data points, we do not spend resources to determine the optimal generation dispatch values, the voltage setpoints, or dual variables before training. Instead, the error factors given in \eqref{EStat} - \eqref{EPrim} will be used to measure the prediction accuracy and train the NN. 

The following loss function is used to modify the shared parameters of the three neural networks:
\begin{align}
&\begin{aligned}
    MAE & = \frac{1}{N_t} \sum_{i=1}^{N_t}  \Lambda_P\underbrace{\lvert \hat{\mathbf G} - \mathbf G\lvert}_{MAE_g} + \Lambda_V\underbrace{\lvert \hat{\mathbf v} - \mathbf v\lvert}_{MAE_v} + \Lambda_L\underbrace{ \lvert \hat{\mathbf L}_{m} - \mathbf L_{m}\lvert}_{MAE_l} \\
    &\qquad + \frac{\Lambda_\epsilon}{N_t + N_c} \sum_{i=1}^{N_t + N_c} \underbrace{\epsilon_{stat} + \epsilon_{comp} + \epsilon_{dual} + \epsilon_{prim}}_{MAE_\epsilon}
    \label{MAE}
\end{aligned}
\end{align}
where $N_t$ is the number of training data points, and $N_c$ is the number of collocation points. The mean absolute error of active and reactive power generation prediction to the actual value is denoted by $MAE_g$. $MAE_v$ and $MAE_l$ indicates the mean absolute error of voltage and dual value prediction, and $MAE_\epsilon$ is the mean absolute value of KKT condition violations given in \eqref{EStat} - \eqref{EPrim}. $\Lambda_P$, $\Lambda_V$, $\Lambda_L$, and $\Lambda_\epsilon$ are the corresponding weights of each loss functions. The physics-informed neural network performance depends significantly on these weights. So, they have to be selected appropriately to reduce either the average error or the maximum constraint violations. 

Since the three NNs are independent, the NN will be trained to minimize the corresponding MAE along with $MAE_\epsilon$. For the collocation points, as discussed earlier, we do not compute the OPF output a priori. As we have not computed the optimal generation dispatch values $\mathbf{G}$, voltage $\mathbf{v}$, and dual variables $\mathbf{L}_m$ for the collocation points, $MAE_g$, $MAE_v$, and $MAE_l$ will be considered zero for these points, and only $MAE_\epsilon$ will be considered in the NN training.
\subsection{Evaluating the Worst Case Performance of the Physics-Informed Neural Network}
The average performance of NN prediction on an unseen test data set is typically used to evaluate and compare different NN architectures. However, in our case, other than the average performance on the test dataset, we will also be using the worst-case generation and line flow constraint violations in the entire input domain to evaluate and improve the performance of the proposed NN training architecture. This section describes the optimization problem used for extracting the worst-case guarantees. The NNs used for predicting the voltages and dual variables are independent of the NN used for predicting the optimal generation set-points. Since they are not required after the training and when the system is ready to be deployed, we can ignore them and instead focus only on the hidden layers used for predicting the generation values. 

\subsubsection{Worst-Case Guarantees for Generation Constraint Violations}
This section discusses the Mixed Integer Linear Programming (MILP) problem formulation used to determine the maximum active and reactive power generation constraint violations, denoted by $v_g$. The maximum constraint violations in the input domain can be formulated as follows:
\begin{subequations}
\begin{gather}
    \underset{\hat{\mathbf G},\mathbf D,\mathbf Z,\mathbf Z^{'},\mathbf y}{\mathrm{max}} v_g \label{WCeq1} \\
    v_g = \mathrm{max}(\mathbf{\hat{G} - \overline{G}}, \mathbf{\underline{G} - \hat{G}},0) \\
    s.t. \text{ } \eqref{NN1}, \eqref{Relu}
\end{gather}
\end{subequations}
where $\overline{G}$ and $\underline{G}$ are the maximum and minimum active and reactive power generation capacity. Since the ReLU activation function, used in \eqref{Relu}, is nonlinear, the mixed-integer reformulation of ReLU activation function proposed in \cite{Andreas} is used as follows:
\begin{subequations}
\begin{align}
    Z^i_k &\leq Z^{' i}_k-Z^{min,i}_k (1-y^i_k) \ &\forall k = 1,..,K \ \forall i = 1,..,N_k \label{RelU1}\\
    Z^i_k &\geq Z^{' i}_k \ &\forall k = 1, ...,K \ \forall i = 1, ...,N_k \label{RelU2}\\
    Z^i_k &\leq Z^{max,i}_k y^i_k \ &\forall k = 1, ...,K \ \forall i = 1, ...,N_k\label{RelU3} \\
    Z^i_k &\geq 0 \ &\forall k = 1, ...,K \ \forall i = 1, ...,N_k \label{RelU4}\\ 
    y_k &\in \{0, 1\}^{N_k} \ &\forall k = 1, ...,K   \label{RelU5} 
\end{align}
\end{subequations}
where $Z^{' i}_k$ and $Z^i_k$ are the inputs and outputs of the ReLU activation function. $Z^{max,i}$ and $Z^{min,i}$ are the maximum and minimum values possible for the respective $Z^{' i}_k$. They should be large enough so that the constraints \eqref{RelU1} and \eqref{RelU3} will not be binding and small enough so that the constraints are not unbounded. $y^i_k$ is a binary variable. If $Z^{' i}_k$ is less than zero, then \eqref{RelU4} will be active. $Z^i_k$ and $y^i_k$ will be zero to satisfy the other constraints. Similarly, if  $Z^{' i}_k$ is greater than zero, then \eqref{RelU2} and \eqref{RelU1} will be active, and $y^i_k$ will be one to satisfy the other constraints. 

\subsubsection{Worst-Case Guarantees for Line Flow Constraint Violations}
Similar to the generation constraints, the line flow limits play an equally significant role in ensuring the proper functioning of the network. However, unlike the generation set-points, we do not directly get as an output the line flow values from the NN prediction. In our effort to formulate the most computationally efficient approach, we observed that simplified convex relaxations such as semidefinite programming for AC Power Flow (AC-PF) were not binding for worst-case guarantees. Therefore, we had to formulate a non-convex power flow problem to obtain the line current flow from the generation set points as follows:
\begin{align}
    \hat{p}_n^g - p_n^d &= \mathbf{\hat{v}}^T_{pf} \mathbf{M}_{p}^n \mathbf{\hat{v}}_{pf}       & \forall n\in N_b \label{PF_NN_p}\\
    \hat{q}_n^g - q_n^d &=\mathbf{\hat{v}}^T_{pf}\mathbf{M}_{q}^n \mathbf{\hat{v}}_{pf}           & \forall n\in N_b \label{PF_NN_q}\\
    \hat{{\ell}}_{mn} &= \mathbf{\hat{v}}^T_{pf} \mathbf{M}_{i_{mn}}  \mathbf{\hat{v}}_{pf} & \forall mn\in N_l \label{I_NN_lim}\\
    \hat{v}^i_{N_{sb}}& = 0 \label{V_NN_sb}
\end{align}
where $\mathbf{\hat{v}}_{pf}$ and $\hat{{\ell}}_{mn}$ are the voltage and square of line flow obtained from the power flow equations.  
The AC-PF equations in \eqref{PF_NN_p}- \eqref{V_NN_sb} are non-convex and quadratic. So, we have to use a Mixed Integer Quadratic Constrained Quadratic Programming (MIQCQP) solver to obtain the worst-case guarantees. The MIQCQP problem can be formulated as follows: 
\begin{subequations}
\begin{gather}
    \underset{\hat{\mathbf G},\mathbf D,\mathbf Z,\mathbf Z^{'},\mathbf y}{\mathrm{max}} v_l \label{WCleq1} \\
    v_l = \mathrm{max}(\hat{\ell}_{mn} - \overline{\ell}^2_{mn},0) \\
    s.t. \eqref{PF_NN_p}-\eqref{V_NN_sb},\eqref{NN1}, \eqref{RelU1}- \eqref{RelU5}
\end{gather}
\end{subequations}
When the  MILP and MIQCQP problems are solved to zero MILP gap, we can ensure that the $v_g$ and $v_l$ values we obtained are the global optima. Thus, we can guarantee that there is no input $\{\mathbf{P}_d, \mathbf{Q}_d \} \in \mathbf{D}$ in the entire input domain, leading to constraint violations larger than the obtained values $v_g$ and $v_l$.
\section{RESULTS \& DISCUSSION}
\subsection{Simulation Setup}
The accuracy and scalability of the proposed physics-informed neural network training architecture are analyzed against a standard NN on four different test systems. The test system characteristics are given in Table I. The network details for case 39, case 118, and case 162 are from the PGLib-OPF network library v19.05 [12], and the network details for case 14 is from \cite{14bus_line}. In all of the test cases, the active and reactive power demand at each node is assumed to be independent of each other. And they are specified to be between 60 to 100\% of their respective maximum loading as follows:
\begin{gather}
    0.6 \text{ }\overline{\mathbf{D}}\leq \mathbf{D} \leq \overline{\mathbf{D}}
\end{gather}
where $\overline{\mathbf{D}}$ denotes the maximum active and reactive power demand. The sum of the maximum loading over all nodes for each system is given in Table \ref{TC}. 

\begin{table}[h]
\centering
\caption{TEST CASE CHARACTERISTICS}
\begin{tabular}{lllllll}
\hline\hline
\multirow{2}{*}{Test Case} & \multirow{2}{*}{$N_{b}$} & \multirow{2}{*}{$N_d$} & \multirow{2}{*}{$N_g$} & \multirow{2}{*}{$N_l$} & \multicolumn{2}{l}{Max loading} \\ \cline{6-7} 
                           &                          &                        &                        &                        & MW             & MVA             \\ \hline \hline
case 14     & 14  & 11  & 5  & 20   & 259  & 276                                                  \\ \hline
case 39     & 39  & 21  & 10  & 46    & 6254   & 6626                                                    \\ \hline
case 118    & 118 & 99  & 19 & 186   & 4242 & 4537                                                      \\ \hline
case 162    & 162 & 113 & 12 & 284   & 7239  & 12005                                                    \\ \hline \hline
\end{tabular}
\label{TC}
\end{table}
Ten thousand sets of random active and reactive power input values were generated using latin hypercube sampling  \cite{hypercube}. From these, 50\% were allotted to the collocation data set, for which we do not have to calculate and provide the OPF setpoints. 20\% of the rest was considered training data points and the remaining 30\% as the unseen test data set. AC-OPF in MATPOWER was used to determine the optimal active and reactive power generation values and voltage setpoints for the input data points in the training and test data sets, and afterwards the KKT conditions given in \eqref{Stat1} -\eqref{Prim} were used to obtain the dual variables.

The properties of both standard and physics-informed neural networks used for the analysis are given in the Table \ref{tab_NN}. We used TensorFlow \cite{tensorflow} with Python for neural network training, and we fixed the number of training epochs to 1,000 and split the data set into 200 batches while training. 

\begin{table}[]
\centering
\caption{Neural Network Properties}
\label{tab_NN}
\begin{tabular}{llllll}
\hline\hline
\multirow{2}{*}{Test Case} & \multicolumn{3}{l}{$N_k$ in Hidden Layer}                       & \multicolumn{2}{l}{Training Time}                                                                         \\ \cline{2-6} 
                           & G                   & V                   & $L_m$               & \begin{tabular}[c]{@{}l@{}}NN\\ (min)\end{tabular} & \begin{tabular}[c]{@{}l@{}}PINN\\ (xNN)\end{tabular} \\ \hline\hline
Case 14                    & 5                   & 10                  & 20                  & 3                                                  & 2                                                   \\ \hline
Case 39                    & \multirow{3}{*}{20} & \multirow{3}{*}{30} & \multirow{3}{*}{50} & 8                                                  & 2.5                                                 \\ \cline{1-1} \cline{5-6} 
Case 118                   &                     &                     &                     & 88                                                 & 3.2                                                 \\ \cline{1-1} \cline{5-6} 
Case 162                   &                     &                     &                     & 105                                                   &  2.6                                                   \\ \hline\hline
\end{tabular}
\end{table}
A High-Performance Computing (HPC) server with an Intel Xeon E5-2650v4 processor and 256 GB RAM was used to train the neural networks. The training time for the standard NN, denoted by NN, and physics-informed neural network, indicated by PINN, as a factor of time taken by the NN is given in Table \ref{tab_NN}. In our experiments the proposed physics-informed neural network took almost three times as much time to train compared to the standard NN. On the other hand, physics-informed neural networks result in significant computational savings when we count in the generation of the training database, as PINNs require considerably less training samples for which both input and output needs to be computed (i.e. we do not need to run an OPF for any collocation point). 

The MILP problem used for solving worst-case guarantees for generation violation and the MIQCQP problem used for solving the worst-case guarantees for line flow violations is formulated in  YALMIP \cite{yalmip} and solved using Gurobi.  The code to reproduce all the simulation results is available online \cite{PINN_Code}.

\subsection{Average Performance over Test Data Set}
This section examines the performance of the physics-informed neural networks versus the standard NNs, considering the most common performance metric for regression NN: the mean absolute error over an unseen test data set. As shown in \eqref{eq:MAE}-\eqref{eq:vdist} and explained below, along with MAE we also define metrics that measure the average constraint violation, sub-optimality, and distance to optimal setpoint:   
\begin{align}
    MAE_T &= \frac{1}{N_T} \sum_{i=1}^{N_T} \lvert \hat{\mathbf{G}} - \mathbf{G}\lvert \label{eq:MAE}\\
    v_g^{avg} &=  \frac{1}{N_T} \sum_{i=1}^{N_T} \mathrm{max}\left(\frac{\hat{\mathbf{G}}- \overline{\mathbf G}}{\overline{\mathbf G}},\frac{\underline{\mathbf G} - \hat{\mathbf{G}}}{\overline{\mathbf G}},0\right) \label{eq:vg}\\
    v_{opt}^{avg}  &= \frac{1}{N_T} \sum_{i=1}^{N_T} \mathbf{c^T}(\hat{\mathbf{G}} - \mathbf{G}) \label{eq:vopt}\\
    v_{dist}^{avg} &=  \frac{1}{N_T} \sum_{i=1}^{N_T} \left( \frac{\lvert \hat{\mathbf{G}} - \mathbf{G} \lvert}{ \overline{\mathbf G} - \underline{\mathbf G}}\right) \label{eq:vdist}
\end{align}
where $MAE_T$ is the mean absolute error with respect to the total generation in test data set, $N_T$ is the test data size, $v_g^{avg}$ is the average active and reactive power generation constraint violation, $v_{opt}$ is the average sub-optimality and $v_{dist}$ is the average distance of predicted value to optimal decision variables $v_{dist}$ in percentage.

During training, both the standard and the physics-informed neural network were optimized to minimize the MAE. However, it was observed that the average performance of the physics-informed neural network in the test data set depends on the hyperparameters used in \eqref{MAE}, while training. So different hyperparameters values were tested, and the combination which produced the least MAE is used to generate the results given in Table \ref{TabAvg}.

\begin{table}[h]
\centering
\caption{Improving Average Performance}
\label{TabAvg}
\begin{tabular}{clllll}
\hline
\hline
\multicolumn{2}{c}{Test Case}        & \begin{tabular}[c]{@{}l@{}}$MAE_T$ \\ (\%)\end{tabular} & \begin{tabular}[c]{@{}l@{}}$v_g^{avg}$ \\ (\%)\end{tabular} & \begin{tabular}[c]{@{}l@{}}$v_{opt}^{avg}$ \\ (\%)\end{tabular} & \begin{tabular}[c]{@{}l@{}}$v_{dist}^{avg}$ \\ (\%)\end{tabular} \\ \hline\hline
\multirow{2}{*}{Case   14} & NN & 0.59                                               & 0.02                                                  & 0.08                                                      & 5.63                                                       \\ \cline{2-6} 
                           & $PINN_{avg}$    & 0.42                                                & 0.00                                                  & 0.05                                                      & 4.43                                                       \\ \hline
\multirow{2}{*}{Case   39} & NN & 0.89                                                & 0.93                                                  & 0.21                                                      & 10.78                                                      \\ \cline{2-6} 
                           & $PINN_{avg}$     & 0.70                                                & 0.88                                                  & 0.08                                                      & 10.38                                                      \\ \hline
\multirow{2}{*}{Case 118}  & NN & 1.07                                                & 1.76                                                  & 0.31                                                      & 11.46                                                      \\ \cline{2-6} 
                           & $PINN_{avg}$     & 0.84                                                & 0.76                                                  & 0.26                                                      & 10.71                                                      \\ \hline
\multirow{2}{*}{Case   162} & NN & 2.22                                                                  & 1.42                                                                      & 0.27                                                                          & 37.81                                                                          \\\cline{2-6} 
                            & $PINN_{avg}$    & 2.00                                                                   & 1.84                                                                     & 0.26                                                                          & 36.72                                                                         
\\ \hline\hline
\end{tabular}
\end{table}

In Table \ref{TabAvg}, NN denotes the standard NN, and $PINN_{avg}$ indicates the results from the physics-informed neural network. 

Incorporating the KKT conditions in the neural network training resulted in a 20\% to 30\% reduction in the MAE on the test dataset. Similarly, a considerable improvement was observed in the average generation constraint violation, suboptimality, and distance from optimality as well. This indicates that we can achieve higher prediction accuracy using a physics-informed neural network with the same or fewer data.

\subsection{Worst Case performance}
While examining the worst-case guarantees for generator violations, we observed no direct correlation between the average performance of the physics-informed neural network in a test data set and the maximum constraint violations. Since worst-case guarantees play a crucial role in the trustworthiness of a NN, similar to the previous case, different hyperparameters were tested to see which one would produce the lowest worst-case generation constraint violations. Then those worst-case violations values, denoted by $PINN$, are compared against (i) the standard NN and (ii) the physics-informed neural network that delivered the least MAE, denoted by $PINN_{avg}$, in Table \ref{TabWC} to access the importance of hyperparameters in the worst-case performance. 

\begin{table}[h]
\centering
\caption{Worst-case Performance }
\renewcommand\tabcolsep{5pt}
\label{TabWC}
\begin{tabular}{llllll}
\hline
\hline
\multicolumn{2}{c}{\multirow{2}{*}{Test Case}} & \multicolumn{2}{c}{$v_g$}                                                                             & \multicolumn{2}{c}{$v_l$}                                                                            \\ \cline{3-6} 
\multicolumn{2}{c}{}                           & \multicolumn{1}{c}{(MW)} & \multicolumn{1}{c}{\begin{tabular}[c]{@{}c@{}}\%  wrt  \\max loading\end{tabular}} & \multicolumn{1}{c}{(MVA)} & \multicolumn{1}{c}{\begin{tabular}[c]{@{}c@{}}\% wrt\\ max loading\end{tabular}} \\ \hline\hline
\multirow{3}{*}{Case 14}       & NN      & 0.09                     & 0.03                                                                            & 38                       & 14.67                                                                          \\ \cline{2-6} 
                               & $PINN_{avg}$      & 0.08                     & 0.03                                                                            & 35                         & 13.51                                                                               \\ \cline{2-6} 
                               & $PINN$           & 0.01                     & 0.01                                                                            & 34                       & 13.13                                                                          \\ \hline
\multirow{3}{*}{Case 39}       & NN       & 246                      & 3.93                                                                            & \multirow{6}{*}{}        & \multirow{6}{*}{}                                                              \\ \cline{2-4}
                               & $PINN_{avg}$      & 184                      & 2.94                                                                            &                          &                                                                                \\ \cline{2-4}
                               & $PINN$          & 129                      & 2.06                                                                            &                          &                                                                                \\ \cline{1-4}
\multirow{3}{*}{Case 118}      & NN       & 266                      & 6.27                                                                            &                          &                                                                                \\ \cline{2-4}
                               & $PINN_{avg}$      & 187                      & 4.41                                                                            &                          &                                                                                \\ \cline{2-4}
                               & $PINN$          & 141                      & 3.32                                                                            &                          &                                                                                \\
                               \cline{1-4}
\multirow{3}{*}{Case 162}      & NN       & 2771                      & 38.27                                                                            &                          &                                                                                \\ \cline{2-4}
                               & $PINN_{avg}$      & 1029                      & 14.21                                                                            &                          &                                                                                \\ \cline{2-4}
                               & $PINN$          & 899                      & 12.42                                                                            &                          &                                                                                \\
                               \hline\hline
\end{tabular}
\end{table}
From Table \ref{TabWC}, we can infer that the physics-informed neural network which produced the lowest MAE also offered a significant reduction in worst-case generation constraint violation compared to the standard NN in all the test cases. This establishes the fact that the physics-informed neural networks offer both better generalization capabilities and lower worst-case guarantees with the same training data compared to a standard NNs. However, an additional 15\% to 30\% reduction in worst-case generation constraint violation is achieved when we tune the hyperparameters of the physics-informed neural networks with the objective to improve specifically the worst-case guarantees instead of the objective to minimize the mean absolute error. This indicates we can further tighten the worst-case guarantees by optimizing the hyperparameters to minimize the worst-case constraint violations.

In `Case 14', along with the generation constraint violations, we were also able to compute the worst-case line flow constraint violations. As  Table~\ref{TabWC} shows, physics-informed neural networks also resulted in lower line flow constraint violations. For Case 39, Case 118, and Case 162, we were unable to compute the worst-case line flow constraint violation since the MIQCQP problem could not be solved to zero optimality gap within 5hr. This highlights the computational challenges associated with the extraction of the worst-case guarantees for the AC-OPF, and in general non-linear, problems. The present work focuses on introducing the first formulations that can extract such guarantees for a non-linear optimization problem. Future work shall focus on addressing the computational issues in order to arrive at scalable algorithms for any system size.

The average performance of $PINN$, (i.e mean absolute error, average generator constraint violations, average suboptimality, and average distance to the optimal setpoint)  is given in Table~\ref{TabAvgWC}. If we compare Table~\ref{TabAvgWC} with Table~\ref{TabAvg}, we observe that in most cases the average performance worsens when we tune the hyperparameters with the objective to reduce the worst-case generation constraint violations instead of the objective to obtain the least mean absolute error. This possibly indicates that there is a trade-off in the performance of the NN between trying to achieve the best average performance and achieving the least worst case violations. However, further analysis is required to ascertain the relationship between good average performance and least worst case violations.
\begin{table}[h]
\centering
\caption{Average Performance }
\label{TabAvgWC}
\begin{tabular}{clllll}
\hline\hline
\multicolumn{2}{c}{Test Case}         & \multicolumn{1}{c}{\begin{tabular}[c]{@{}c@{}}$MAE_T$ \\ (\%)\end{tabular}} & \multicolumn{1}{c}{\begin{tabular}[c]{@{}c@{}}$v_g^{avg}$ \\ (\%)\end{tabular}} & \multicolumn{1}{c}{\begin{tabular}[c]{@{}c@{}}$v_{opt}^{avg}$ \\ (\%)\end{tabular}} & \multicolumn{1}{c}{\begin{tabular}[c]{@{}c@{}}$v_{dist}^{avg}$ \\ (\%)\end{tabular}} \\ \hline\hline
{Case 14} &$PINN$    & 0.60                                                                    & 0.07                                                                      & 0.06                                                                          & 5.75                                                                           \\ \hline
{Case 39} &$PINN$  & 2.17                                                                    & 1.19                                                                      & 0.40                                                                          & 12.73                                                                          \\ \hline
{Case 118} &$PINN$  & 1.04                                                                   & 1.80                                                                      & 0.18                                                                          & 10.31                                                                          \\ \hline
{Case   162} &$PINN$  & 2.27                                                                   & 0.97                                                                      & 0.25                                                                          & 39.13                                                                          \\ \hline\hline
\end{tabular}
\end{table}

\subsection{Input Domain Reduction}
Ref.~\cite{Andreas} has shown that for linear optimization programs, the worst-case constraint violation of the NN can be reduced by training the NN on a broader input domain than the one on which it will be deployed. We explore how this approach performs for non-linear programs, both for standard NNs and PINNs. For both neural network types, the input domain for the test set was symmetrically reduced by $\delta$ as follows:
\begin{align}
    (0.6+\delta)\overline{\mathbf{D}}\leq \mathbf{D} \leq (1- \delta)\overline{\mathbf{D}}
\end{align}

\begin{figure}[h]
\centerline{\includegraphics[scale=0.47]{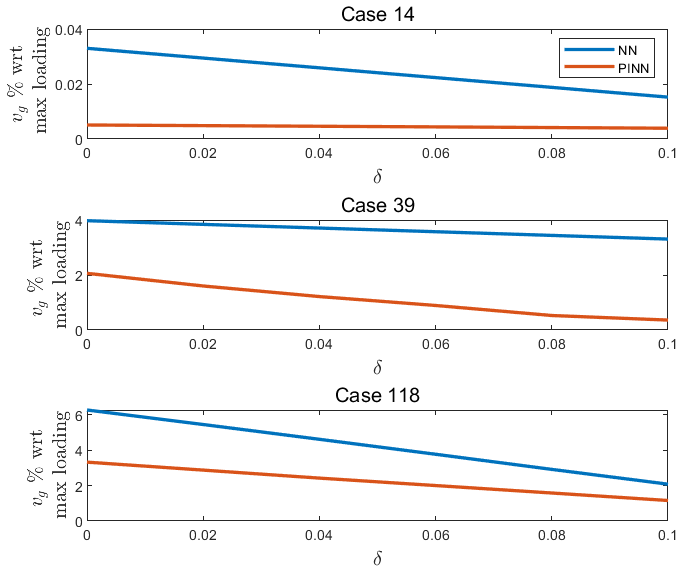}}
\caption{Input Domain Reduction}
\label{Domain}
\end{figure}

The resulting worst-case generation constraint violation with respect to the maximum active power loading is given in Fig.~\ref{Domain}. 

We observe that similar to linear programs, for the AC-OPF as well input domain reduction of test sets improves the worst-case guarantees considerably, especially for the standard NNs.  We notice that the reduction of the violations is significantly higher for standard NNs compared with the Physics-Informed Neural Networks (PINNs), although across the whole input domain reduction the PINNs exhibit always lower worst-case violations. This implies that the physics-informed neural network will only require a slightly larger input domain of the training data set to maintain the worst-case constraint violations below a certain threshold, while standard NNs would require significantly larger input domains to achieve a similarly low level of worst-case constraint violation, if at all possible.

\section{Conclusion}
This paper offers two key contributions. First, we introduced a framework to incorporate the physical equations of the AC-OPF inside the neural network training. Additionally, we show that we can improve the NN prediction accuracy by using physics-informed neural networks while using considerably fewer input data points. Second, we propose a method to extract and minimize the worst-case generation and line constraints violations for the AC-OPF problem. In the future, the work will be extended to include a multilevel optimization algorithm to determine the key physics-informed neural network hyperparameters which minimize worst-case constraint violations.

\bibliographystyle{IEEEtran}
\bibliography{references}

\end{document}